# Relative contributions of cross-even and cross-odd parts to the spin dependent elastic hadron scattering amplitudes at high energies

O.V. Selyugin(https://orcid.org/0000-0001-8933-2834)[a]

[a] *Joint Institut for Nuclear Researches, Dubna, 141490 Russia*

*e-mail: selugin@theor.jinr.ru*

**Abstract**—The form and energy dependence of different terms of the cross-even and cross-odd parts of the elastic nucleon-nucleon scattering amplitude is determined in the framework of the High Energy Generalize Structure (HEGS) model. In the framework of the HEGS model, using the electromagnetic and gravitomagnetic form factors, the differential cross sections in the Coulomb Nuclear Interference (CNI) region and at large momentum transfer are described well in a wide energy region simultaneously. It is shown that the cross-even part includes the soft pomeron growing like $\mathrm{Log}(s)^2$ and an additional term with a large slope and with $\mathrm{Log}(s)$ growth. The cross-odd part includes the maximal odderon term and an additional oscillation term $\mathrm{Log}(s)$ growth. It is shown that both additional terms are proportional to charge distributions, but the maximal odderon term is proportional to matter distributions.

## INTRODUCTION

The development of our understanding of the hadron structure and interactions is mostly based on new experimental data, especially those for the elastic proton-proton scattering obtained at the LHC by the TOTEM-SMS and ATLAS Collaborations. In the case of long-range potentials, one of the key challenges is related to the structure of the scattering amplitude at small angles. The main properties of the hadron scattering amplitude such as analyticity, unitarity, and crossing-symmetry are tightly connected with the first principles of quantum field theory [1-3] and the concept of the scattering amplitude as a unified analytic function of its kinematic variables [4].

The determination of the structure of the elastic hadron scattering amplitude at super-high energies and small momentum transfer $-t$ give us a possibility to establish a correspondence between experimental knowledge and basic asymptotic theorems. This connection gives information about hadron interaction at large distances where the perturbative QCD does not work [5], and a new theory can be developed. Analysis of new experimental data requires improve method of the such analysis.

In this paper, an example of the successful application of a new technique is demonstrated. The study is dedicated to a more accurate assessment of the cross-even and cross-odd parts of the elastic hadron scattering amplitude.

# MODEL DESCRIPTIONS OF DIFFERENTIAL CROSS SECTIONS IN WIDE ENERGY AND MOMENTUM TRANSFER REGIONS

The differential cross sections of elastic nucleon scattering are described by five spin-dependen parts

$$\frac{d\sigma}{dt} = \frac{2\pi}{s^2}[\Phi_1^2(s,t) + \Phi_2^2(s,t) + \Phi_3^2(s,t) + \Phi_4^2(s,t) + 4\Phi_5^2(s,t)]. \quad (1)$$

Every spiral amplitude includes the hadronic and electromagnetic parts $\Phi_i(s,t) = F_i^h(s,t) + F_i^{em}(t)e^{\varphi(s,t)}$, where $F_i^h(s,t)$ is determined by strong and $F_i^{em}(t)$ electromagnetic interactions, and $\phi(s,t)$ is the phase factor determined by the Coulomb-hadron interference [6-8]. Each amplitude includes the corresponding nucleon form factor. These form factors in the HEGS model are calculated from the same generalized parton distributions which were obtained in [9-11]. The electromagnetic amplitudes are taken into account in the standard form [12]. In the model the eikonal approximation is used. Therefore, the Born scattering amplitudes are first determined and the eikonal phase are calculated by numerical

$$\chi_0(s,b) = 2\pi \int_0^\infty q\, e^{i\vec{q}\vec{b}}\, J_0(bq)\, F_0^B(s,q)\, dq \quad (4)$$

where the $J_0(x)$ is the Bessel function of zero order. The phase is strictly connected with the hadron interaction potential. Then, using the eikonal form, which can be obtained from the nonlinear equation [13], the full scattering amplitude in the tree approach, is obtained. In the numerical calculation, the integral form of the Bessel function was used for the spin-non-flip amplitudes.

In general, it is believed that the elastic scattering amplitude is proportional to the electromagnetic form factors of the interaction hadrons. Strong interactions can be connected with a more complicated structure, for example, with the mass distribution in the hadron. As a result, many successful models, for example, the Bourrely-Soffer-Wu model [14], take into account the form factors as some fitting functions which reflect the electromagnetic and mass structure of the hadrons. A more complicated picture of the hadron structure appears with introducing the general (or non-forward) parton distributions - (GPDs) [15-16], (see, for example, the review [17]), including both spin-independent and spin-dependent distributions, for example, such as $H(x,\xi,t), E(x,\xi,t)$. The GPDs can be determined as the Furrier representation of the non-local matrix element [16].

The different moments of GPDs allow us to calculate the different hadron form factors such as the Compton form factors $(R_V(t), R_T(t), R_A(t))$, the electromagnetic form factors $(F_1(t), F_2(t))$ and the gravitomagnetic form factors $(A(t), B(t))$ [11], which reflect the mass distribution of the hadron. The GPDs include part of the spin dependent and spin-independent parton distributions, which depend on the Biorken variable $x$, and an additional dependence on the momentum transfer $t$ and an additional variable $\xi$ (the parameter of the asymmetry). According to the sum rules [15], the electromagnetic form factors of the hadrons can be obtained as a sum of the quark distributions. The non-factorization dependence of the momentum transfer of the leading GPDs $H(x,\xi,t), E(x,\xi,t)$ model was proposed in [11], in the framework of which good descriptions of the proton and neutron form factors were obtained. Further development of the model requires the analysis [16] of the different parton distributions (PDFs) obtained from the experimental data on deep inelastic scattering (PDFs). The nucleon form factors are calculated numerically using the representation of the obtained GPDs distribution [18].

The separate parts of the Born scattering amplitude are determined as

$$F_1^B(s,t) = h_2 G_{em}^2(t)\,(\hat{s})^{\Delta_1}\,e^{\alpha_1 t \ln(\hat{s})}; \qquad F_3^B(s,t) = h_3 G_A(t)^2\,(\hat{s})^{\Delta_1}\,e^{\alpha_1/4 t \ln(\hat{s})};$$

$$F^B(\hat{s},t) = F_1^B(\hat{s},t)(1+R_1/\sqrt{\hat{s}})] + F_3^B(\hat{s},t)(1+R_2/\sqrt{\hat{s}})] + F_{odd}^B(s,t);$$

$$F_{Odd}^B(s,t) = h_{Odd}\,G_A(t)^2\,(\hat{s})^{\Delta_1}\,\frac{t}{1-r_o^2 t}e^{\alpha'/4 t \ln(\hat{s})};$$

The form factors $F_1(t)$ and $A(t)$ are determined by the first and second moments of GPDs, respectively, and reflect the charge and matter distributions. The model takes into account the Odderon contribution. So the full Born term of the scattering amplitude is

$$F_h^{Born}(s,t) = h_1 F_1^2(t) F_a(s,t)\left(1+\frac{r_1}{\sqrt{\hat{s}}}\right) + h_2 A^2(t) F_b(s,t) \pm h_{odd} A^2(t) F_b(s,t)\left(1+\frac{r_2}{\sqrt{\hat{s}}}\right) \qquad (10$$

The important property of the model consists in that the real part of the scattering amplitude is determined only by the complex cross symmetric form of energy $\hat{s} = se^{-i\pi/2}$. No other artificial functions or any parameters impact the form or s and t dependence of the phase of the scattering amplitude. Note that the role of the real part is especially important at low momentum transfer (where the differential cross sections are determined by the Coulomb-hadron interference effects) and in the region of the diffraction minimum (where the imaginary part of the scattering amplitude has the zero, and the size of the diffraction minimum is determined by the real part of the scattering amplitude and the small contribution to the CNI term. As a result, a good description of the differential cross section is obtained in a wide energy region (as an example see Fig.1 (a,b) ).

## NEW EFFECTS ARE DISCOVERED IN THE DIFFRACTON SCATTERING

## AT SMALL ANGLES

In the fitting procedure of the experimental data, only statistical errors were taken into account. As the systematic errors are mostly determined by indefiniteness of luminosity, they are taken into account as an additional normalization coefficient. This method essentially decreases the space of a possible form of the scattering amplitude. This allows us to find the appearance of some small effects at 13 TeV experimental data for the first time [19-21].

Our further researches with taking into account a wider range of experimental data confirm such new effects. We determined the new anomalous term with a large slope as

$$F_{an}(s,t) = ih_{an} G_{em}^2(t) Ln(\hat{s}/k) \, e^{-\alpha_{an}(|t|+(2t)^2)\ln(\hat{s})};$$

It is proportional to electromagnetics form factors and the analysis of the experimental data above 500 GeV gives the sizes of the constant $h_{an}$ (see Table 1). When such a term is analyzed in a wide energy region with inclusion of experimental data on both proton-proton and proton-antiproton elastic scattering, the constant $h_{an}$ has a maximal value and a minimal error.
Our method helps us to find a small oscillation effect in the differential cross section at small momentum transfer. Such oscillation can be determined by an additional oscillation term in the scattering amplitude. It is determined as

$$F_{osc}^{ad}(s,t) = \pm h_{osc} J_1[\tau]/\tau; \quad \tau = \pi(\varphi_0 - t)/t_0;$$

The effect of oscillation can be seen in the difference between experimental data and theoretical calculations without such an oscillation term, which is divided by the smooth curve and compared it with this difference, but instead of experimental data it one should take the model calculating the differential cross sections including the oscillation term.
As a result of the analysis of proton-proton and proton antiproton elastic scattering in the energy region from 6 GeV up to 13 TeV it was shown that such an oscillation term has a small energy dependence and has cross-odd properties. It was obtained in the form

$$F_{osc}^{ad}(s,t) = ih_{osc} G_{em}^2(t) \, J_1[\tau]/\tau [Ln(\hat{s}/k) + h_{c1}](1 + ih_{c2});$$

where $G^2_{em}(t)$ is the electromagnetic form factor. This oscillation term has small energy independent part, which is proportional to the constant $h_{c1}$ and the energy dependent part proportional to the constant $h_{osc}$.
In Fig.2, the ratio of the cross-odd part of the scattering amplitude of elastic proton-proton scattering to its cross even part at 13 TeV and at 9.26 GeV is presented. It is seen that this ratio is small at 13 TeV, except the position of the diffraction minimum. At low energy, this ratio is small at small t but increases essentially in the region of large momentum transfer.

## CONCLUSIONS

The paper studies the elastic nucleon-nucleon scattering in a wide energy region (from 6 GeV up to 13 TeV in the framework of the high energy generalized structure model. The relative contributions of the possible different part of the scattering amplitude were especially analyzed. It was found that the new anomalous term with a large slope has the logarithmic energy dependence and has the cross even properties. Hence, it is part of the pomeron amplitude and is also proportional to charge distributions. Our analysis of the contribution of the so-called hard pomeron with a large intercept does not show a visible contribution of this term. This corresponds to the final Landshofft conclusion.
The second additional term, which represents the oscillation properties of the scattering amplitude at small momentum transfer, has the cross-odd properties. Hence, it belongs to the odderon contribution in the scattering amplitude. This term has a small energy independent part and energy dependent part with Log(s) growth. In the model, the main pomeron and odderon amplitudes have the same intercept. In this sense, we have the maximal Odderon introduced by B. Nicolescu. However, the odderon amplitude has a special kinematic factor and does not give a visible contribution at zero momentum transfer. It corresponds to the Landshofft picture of the odderon.

## FUNDING

This research was carried out at the expense of the grant of the Russian Science Foundation No. 23-22-00123, https://rscf.ru/project/23-22-00123 /.

## CONFLICT OF INTEREST

The authors declare that they have no conflicts of interest.

**Table 1.** The determination of the size of the anomalous term based on experimental data in different energy regions

| √s = 13 TeV, (pp) | √s = 13 TeV – 7 TeV, (pp) | √s > 500 GeV, (pp and p$\bar{p}$) |
|---|---|---|
| $h_{an}^a = 1.75 \pm 0.05 \ GeV^{-2}$ | $h_{an}^b = 1.54 \pm 0.08 \ GeV^{-2}$ | $h_{an}^a = 2.27 \pm 0.05 \ GeV^{-2}$ |

FIGURE CAPTIONS

**Fig. 1.** (a) Differential cross section of elastic proton proton scattering at 13 TeV (line –our model calculations, dotes – the data of the TOTEM (SMS) Collaboration); (b) ) Differential cross section of elastic proton proton scattering at 9.26 GeV (line –our model calculations, dotes – the data of [20].

**Fig. 2:** The ratio of the cross-odd part of the scattering amplitude of elastic proton-proton scattering to its cross even part (a) at 13 TeV;   (b) at 9.26 GeV.

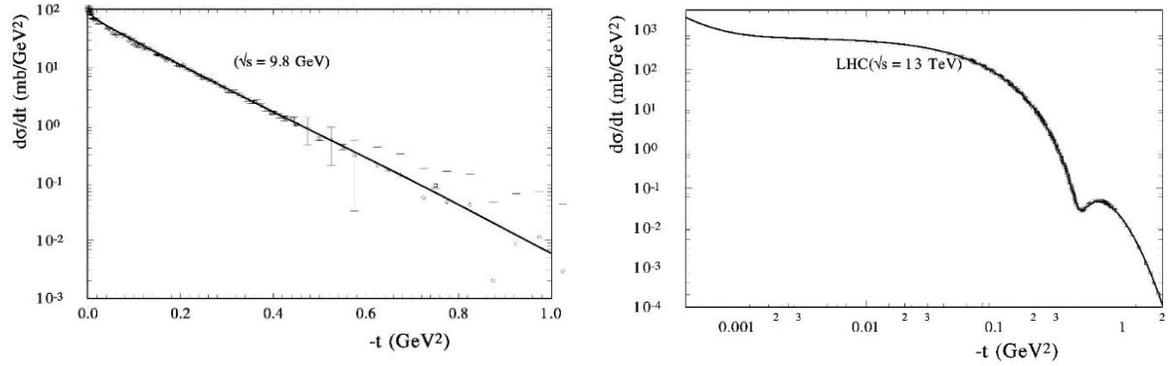

Fig. 1 (a),(b).

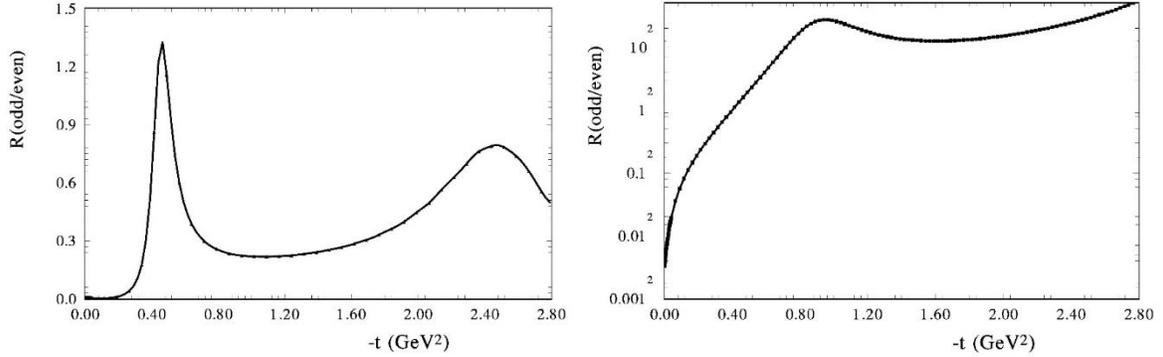

Fig. 2 (a,b).